\documentclass[twocolumn,english,aps,pra,showpacs]{revtex4}
\usepackage{comment,amsmath,graphicx,amssymb,epsfig,babel,dsfont,color,subfigure}

\newcommand{\rd}{{\rm d}}

\newcommand{\ri}{{\rm i}}
\newcommand{\re}{{\rm e}}
\newcommand{\ro}{{\rm o}}
\newcommand{\rs}{{\rm s}}
\newcommand{\rp}{{\rm p}}

\begin{document}

\title{Long-Range Dipole-Dipole Interaction and Anomalous F\"{o}rster Energy Transfer across Hyperbolic Meta Material}

\author{S.-A. Biehs$^*$}

\affiliation{Institut f\"{u}r Physik, Carl von Ossietzky Universit\"{a}t, D-26111 Oldenburg, Germany\\
$^*$Corresponding author: s.age.biehs@uni-oldenburg.de}

\author{Vinod M. Menon}

\affiliation{Department of Physics, City College of New York, 160 Convent Ave New York, NY 10031, USA}

\author{G. S. Agarwal}

\affiliation{Department of Physics, Oklahoma State University, Stillwater, Oklahoma 74078, USA}

\date{\today}


\begin{abstract}
We study radiative energy transfer between a donor-acceptor pair across a hyperbolic metamaterial
slab. We show that similar to a perfect lens a hyperbolic lens allows for giant energy transfer rates. For a realistic
realization of a hyperbolic multilayer metamaterial we find an enhancement of up to three orders of 
magnitude with respect to the transfer rates across a plasmonic silver film of the same size especially
for frequencies which coincide with the epsilon-near zero and the epsilon-near pole frequencies. Furthermore,
we compare exact results based on the S-matrix method with results obtained from effective medium theory.
Our finding of very large dipole-dipole interaction at distances of the order of a wavelength has important 
consequences for producing radiative heat transfer, quantum entanglement etc.
\end{abstract}

\maketitle

%%%%%%%%%%%%%%%%%%%%%%%%%%%%%%%%%%%%%%%%%%%%%%%%%%%%%%%%%%%%%%%%%%%%%%%%%%%%%%%%%%%%
%
% Introduction 
%
%%%%%%%%%%%%%%%%%%%%%%%%%%%%%%%%%%%%%%%%%%%%%%%%%%%%%%%%%%%%%%%%%%%%%%%%%%%%%%%%%%%%

\section{Introduction}

Dipole-dipole interactions are at the heart of many fundamental interactions such as 
van der Waals forces and vacuum friction~\cite{MilonniBook,Volokitin2007}, F\"{o}rster (radiative) energy transfer (FRET)~\cite{Forster,DungEtAl2002}, radiative heat transfer~\cite{Volokitin2007,BiehsAgarwal2013b}, 
quantum information protocols like the realization of CNOT gates~\cite{Nielsen,Bouchoule2002,Isenhower2010}, pairwise excitation of 
atoms~\cite{VaradaGSA1992,HaakhEtAl2015,HettichEtAl2002}, and Rydberg blockade~\cite{SaffmanEtAl,GilletEtAl}.
Clearly large number of problems in physics and chemistry require very significant dipole-dipole interaction 
at distances which are not much smaller than a wavelength.
The development of plasmonic and metamaterial platforms can considerably enhance these
fundamental dipole-dipole interactions. In an early work~\cite{GSASub} it was shown that 
the dipole-dipole interaction~\cite{GSASub,ElGanainy2013} can be quite significant even at distances bigger than 
microns if one utilizes whispering gallery modes of a sphere. More recently such systems have 
been revisited for their remarkable quantum features like squeezing~\cite{HaakhEtAl2015}. Further it
was shown that the energy transfer across plasmonic metal films can be enhanced~\cite{AndrewBarnes2004}
and that the long range plasmons allow for long-range plasmon assisted energy transfer 
between atoms placed on plasmonic structures such as graphene~\cite{Velizhanin2012, AgarwalBiehs2013,KaranikolasEtAl2016} 
and metals~\cite{BiehsAgarwal2013,Poudel2015,LiEtAl2015,CanoEtAl2010,BouchetEtAl2016}.

More elaborate structures possessing more features
than simple single-layer plasmonic structures are for ex-
ample hyperbolic metamaterials (HMM)~\cite{HuChui2002,Smith2003}, which
exhibit a broadband enhanced LDOS~\cite{SmolyaninovNarimanov2010} allowing for
broadband enhanced spontaneous emission~\cite{ZubinEtAl2012,PoddubnyEtAl2011,ChebykinEtAl2012,KidwaiEtAl2012,IorshEtAl2012,PotemkinEtAl2012,KidwaiEtAl2011,OrlovEtAl2011,KimEtAl2012,KrishnamorthyEtAl2012,GalfskyEtAl2015,WangEtAl2015}, hy-
perbolic lensing~\cite{JacobEtAl2006,FengAndElson2006,CegliaEtAl2014}, negative refraction~\cite{SmithEtAl2004,HoffmanEtAl2007} and
broadband enhanced thermal emission~\cite{Nefedov2011, Biehs2012,GuoEtAl2012,Biehs2,ShiEtAl2015,Biehs2015}, for instance. 
These HMM can be artificially fabricated by a periodic layout of sub-wavelength metal and dielectric components 
but they also exist in nature~\cite{ThompsonEtAl1998,CaldwellEtAl2014,EsslingerEtAl2014,NarimanovKildishev2015,KorzebEtAl2015}. 
Here we concentrate on artificially fabricated HMM since the choice of fabrication parameters makes it possible to
control the epsilon-near zero (ENZ) and the epsilon-near pole (ENP) frequencies. At these frequencies the HMM
can show extra-ordinary features as enhanced superradiance~\cite{Fleury2013} and supercoupling~\cite{Silvereinha2006,EdwardsEtAl2008}. 

\begin{figure}[Hhbt]
  \epsfig{file = 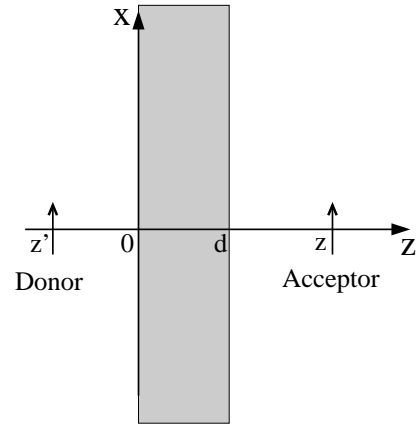, width = 0.3\textwidth}
  \caption{\label{Fig:geometry} Sketch of the considered geometry. The metamaterial slab at the moment is an 
           isotropic one with $\epsilon$ and $\mu$ as the dielectric and magnetic permeabilities.}
\end{figure}

%%%%%%%%%%%%%%%%%%%%%%%%%%%%%%%%%%%%%%%%%%%%%%%%%%%%%%%%%%%%%%%%%%%%%%%%%%%%%%%%%%%%
%
% Letter Summary
%
%%%%%%%%%%%%%%%%%%%%%%%%%%%%%%%%%%%%%%%%%%%%%%%%%%%%%%%%%%%%%%%%%%%%%%%%%%%%%%%%%%%%

The aim of our work is to study FRET across a hyperbolic multilayer metamaterial (see Fig.~\ref{Fig:geometry}). 
We will demonstrate that similar to the perfect lens effect~\cite{Pendry2000}, the dipole-dipole interaction
across a HMM can be enhanced by orders of magnitude compared to the pure plasmonic enhancement
by a thin plasmonic metal layer.

The organization of our paper is as follows: In Sec.~II we introduce the elementary relations for
FRET between a donor and an acceptor which are separated by an intermediate
slab. We discuss the possibility of a perfect dipole-dipole interaction which is studied in Sec.~III 
analytically and numerically for a thin silver film. Then, in Sec.~IV, we discuss
the anomalous transmission across a hyperbolic metamaterial as an alternative for realizing perfect
dipole-dipole interaction. Finally, in Sec.~V we study FRET across a concrete hyperbolic structure 
using exact S-matrix calculations. The exact numerical results are compared with 
the approximative results of effective medium theory.

%%%%%%%%%%%%%%%%%%%%%%%%%%%%%%%%%%%%%%%%%%%%%%%%%%%%%%%%%%%%%%%%%%%%%%%%%%%%%%%%%%%%
%
% Theoretical Model: Förster Energy Transfer 
%
%%%%%%%%%%%%%%%%%%%%%%%%%%%%%%%%%%%%%%%%%%%%%%%%%%%%%%%%%%%%%%%%%%%%%%%%%%%%%%%%%%%%
\section{Possibility of Perfect Dipole-Dipole Interaction}

The F\"{o}rster energy transfer rate between an donor-acceptor pair as sketched in Fig.~\ref{Fig:geometry} is well-known
and can be written in terms of the dyadic Green's function $\mathds{G}$ as~\cite{DungEtAl2002}
\begin{equation}
  \Gamma = \int \!\!\! \rd \omega\, \sigma^{\rm abs}(\omega) T(\omega) \sigma^{\rm em}(\omega)
\end{equation}
where $\sigma^{\rm abs}$ and $\sigma^{\rm em}$ are the absorption and emission spectra of the acceptor and donor and
\begin{equation}
  T(\omega) = \frac{2 \pi}{\hbar^2} \biggl( \frac{\omega^2}{\epsilon_0 c} \biggr)^2 |\mathbf{d}_{\rm D}|^2 |\mathbf{d}_{\rm A}|^2 |\mathbf{e}_{\rm A} \cdot \mathds{G} \cdot \mathbf{e}_{\rm D}|^2
\end{equation}
introducing the dipole-transition matrix elements $\mathbf{d}_{\rm D} = |\mathbf{d}_{\rm D}| \mathbf{e}_{\rm D}$ and $\mathbf{d}_{\rm A} = |\mathbf{d}_{\rm A}| \mathbf{e}_{\rm A}$ of the donor and acceptor.

Making a plane wave expansion of the dyadic Green's function, we obtain
\begin{equation}
  \mathds{G}(\mathbf{r,r'}) =  \int \!\!\frac{\rd^2 \kappa}{(2 \pi)^2} \, \re^{\ri \boldsymbol{\kappa}\cdot\mathbf{X}} \mathds{G}(\boldsymbol{\kappa},z)
\label{Eq:GreenIntegral}
\end{equation}
where $\boldsymbol{\kappa} = (k_x,k_y)^t$, $\mathbf{X} = (x-x',y-y')^t$, and 
\begin{equation}
   \mathds{G} (\boldsymbol{\kappa},z) = \frac{\ri \re^{\ri\gamma_v (z - z')}}{2 \gamma_v} \biggl[  t_\rs \mathbf{a}_\rs^+ \otimes \mathbf{a}_\rs^+ + t_\rp \mathbf{a}_\rp^+ \otimes \mathbf{a}_\rp^+ \biggr]
\label{Eq:GreensTensor}
\end{equation}
introducing the vacuum wavevector in z-direction $\gamma_v = \sqrt{k_v^2 - \kappa^2}$ and the vacuum wavenumber $k_v = \omega/c$. Note that the expression for the Green's function contains the contribution of the propagating ($\kappa < k_v$) and
evanescent ($\kappa > k_v$) waves. The evanescent wavas are exponentially decaying in z-direction so that during propagation
these contributions are lost unless the medium somehow can reverse the decay of such waves. 
Here $t_\rs$ and $t_\rp$ are the transmission coefficients of the s- and p polarization 
and $\mathbf{a}_{\rs,\rp}^+$ are the polarization vectors defined by
\begin{equation}
  \mathbf{a}_\rs^+ = \frac{1}{\kappa} \begin{pmatrix} k_y \\ - k_x \\ 0 \end{pmatrix} 
   \quad \text{and} \quad
  \mathbf{a}_\rp^+ = \frac{1}{\kappa k_v}\begin{pmatrix} - k_x \gamma_v \\ - k_y \gamma_v \\ \kappa^2 \end{pmatrix}.
\end{equation}
The transmission coefficients for a metamaterial film can be expressed as~\cite{Pendry2000,HuChui2002}
\begin{align}
  t_\rs &= \frac{4 \mu \gamma \gamma_v \re^{\ri (\gamma - \gamma_v) d}}{(\gamma + \mu \gamma_v)^2 - (\gamma - \mu \gamma_v)^2 \re^{2 \ri \gamma d}}, \\
  t_\rp &= \frac{4 \epsilon \gamma \gamma_v \re^{\ri (\gamma - \gamma_v) d}}{(\gamma + \epsilon \gamma_v)^2 - (\gamma - \epsilon \gamma_v)^2 \re^{2 \ri \gamma d}}, \label{Eq:tpMM}
\end{align}
where the wavevector component inside the medium in z-direction is given 
by $\gamma = \sqrt{\epsilon \mu \omega^2/c^2 - \kappa^2}$.
Note that the dipole-dipole interaction in free space (i.e.\ if we replace 
the film by vacuum) can be obtained from the Green's tensor in Eq.~(\ref{Eq:GreensTensor})
by setting the transmission coefficients to one $t_\rs = t_\rp = 1$ (which is the result for $\epsilon = \mu = 1$), i.e.\ 
\begin{equation}
   \mathds{G}^{\rm (vac)} (\boldsymbol{\kappa},z) = \frac{\ri \re^{\ri\gamma_v (z - z')}}{2 \gamma_v} \biggl[ \mathbf{a}_\rs^+ \otimes \mathbf{a}_\rs^+ + \mathbf{a}_\rp^+ \otimes \mathbf{a}_\rp^+ \biggr]
\label{Eq:GreenFreeSpace}
\end{equation}
In this case the dipole-dipole interaction would become infinitely large if the donor and the acceptor would be placed 
at the same position, that means if $z = z'$. From a mathematical point of view this is so because the exponential 
prefactor equals one for $z = z'$ so that the $\kappa$-integral in Eq.~(\ref{Eq:GreenIntegral}) would diverge due to
the fact that an infinite number of evanescent waves with $\kappa > k_v$ would contribute to the energy transfer. 
On the other hand, it is well known that the dipole-dipole interaction between the donor and acceptor is proportional to $1/|\mathbf{r} - \mathbf{r}'|^3$ which diverges for $z \rightarrow z'$, since we here have $x = x' = 0$ and $y = y' = 0$ as shown in the Fig.~\ref{Fig:geometry}.

%%%%%%%%%%%%%%%%%%%%%%%%%%%%%%%%%%%%%%%%%%%%%%%%%%%%%%%%%%%%%%%%%%%%%%%%%%%%%%%%%%%%
%
% Theoretical Model ideal left-handed material
%
%%%%%%%%%%%%%%%%%%%%%%%%%%%%%%%%%%%%%%%%%%%%%%%%%%%%%%%%%%%%%%%%%%%%%%%%%%%%%%%%%%%%
%\section{Ideal left-handed material}

\begin{figure}[Hhbt]
  \epsfig{file = 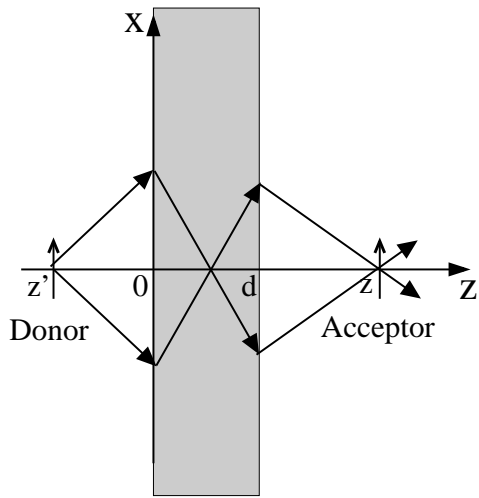, width = 0.35\textwidth}
  \caption{\label{Fig:geometry2} Sketch of the perfect lensing effect.}
\end{figure}

For an ideal left-handed material exhibiting negative refraction the permittivity and permeability are 
given by $\epsilon = \mu = -1$. In such a material evanescent plane waves are amplified as shown by 
Pendry when he introduced the concept of a perfect lens~\cite{Pendry2000}. For the here defined 
transmission coefficients we obtain for $\epsilon = \mu = -1$
\begin{equation}
   t_\rs = t_\rp = \exp(-2 \ri \gamma_v d)
\end{equation}
showing clearly the feature of amplification of evanscent waves inside the ideal left-handed material. 
Note, that this exponential amplification stems from the exponential in the denominator of the transmission 
coefficients. This is just the same expression derived by Pendry showing that for the perfect lens
{\itshape both propagating and evanescent waves contribute to the resolution of the image}~\cite{Pendry2000}
which leads to the perfect lensing effect depicted in Fig.~\ref{Fig:geometry2}.

Inserting the expression of the transmission coefficients in the Green's tensor describing the dipole-dipole interaction 
gives
\begin{equation}
   \mathds{G} (\boldsymbol{\kappa},z) = \frac{\ri \re^{\ri\gamma_v (z - z' - 2d)}}{2 \gamma_v} \biggl[\mathbf{a}_\rs^+ \otimes \mathbf{a}_\rs^+ + \mathbf{a}_\rp^+ \otimes \mathbf{a}_\rp^+ \biggr].
\end{equation}
This Green's tensor is the same as expression (\ref{Eq:GreenFreeSpace}) of the vacuum Green's tensor with the 
important difference that now the argument of the exponential prefactor is different. As in the case of interaction 
in free space we have an infinite large energy transfer if the exponential equals one leading to the condition 
that $z - z' = 2d$. Hence by placing the donor-acceptor pair such that this condition is fullfilled corresponds 
to a dipole-dipole interaction for two dipoles at the same position leading to an infinitely large energy transfer 
rate. This is so, because for $z - z' = 2d$ all evanescent waves ($\kappa > k_v$) get focused. We have thus shown 
that a pefect negative material with zero losses can yield perfect dipole-dipole interaction and perfect energy transfer 
which is limited only by the donor and acceptor line shapes.
% The same effect can be achieved with a left-handed uni-axial material 
%as shown by Hu and Chui~\cite{HuChui2002}. 

\section{Silver Film as Metamaterial}

As pointed out by Pendry~\cite{Pendry2000} for observing the perfect lensing effect it suffices to consider
a thin silver film. In the quasistatic-limit such a thin silver film mimicks a perfect lens for the p-polarized
waves at a frequency $\omega_{\rm pl}$ where $\epsilon'(\omega_{\rm pl}) = -1$, which coincides with the surface 
plasmon resonance frequency of a single metal interface. To see this we take the quasi-static limit ($\kappa \rightarrow \infty$)
of the transmission coefficient $t_\rp$ in Eq.~(\ref{Eq:tpMM}) with $\mu = 1$. We obtain
\begin{equation}
  t_\rp \rightarrow \frac{4 \epsilon}{(\epsilon + 1)^2 - (\epsilon - 1)^2 \re^{2 \ri \gamma d}}.
\end{equation}
The exponential in the numerator vanishes because in the quasi-static limit $\gamma_v \approx \gamma \approx \ri \kappa$.
If we now insert
\begin{equation}
   \epsilon = -1 + \ri \epsilon'',
\end{equation}
then we arrive at
\begin{equation}
  t_\rp \rightarrow \frac{4 (\ri \epsilon'' - 1)}{{\epsilon''}^2 (\re^{2 \ri \gamma d} - 1) + 4 (\ri \epsilon'' - 1)\re^{2 \ri \gamma d}}.
\end{equation}
It is now easy to see that for vanishing losses $\epsilon'' \rightarrow 0$ we again have ($\gamma_v \approx \gamma \approx \ri \kappa$) 
\begin{equation}
  t_\rp = \exp(-2 \ri \gamma_v d).
\end{equation}
Proving that a silver film can mimick a perfect lens. The only drawback 
is that in metals the losses are not negligible so that the perfect lens effect does not persist in 
this case. Nonetheless even with losses one can expect to find large  F\"{o}rster 
energy transfer due to the fact that the donor and acceptor can couple to the surface waves 
inside the silver film.

\begin{figure}[Hhbt]
  \subfigure[\, Ag, $F_z$]{\epsfig{file = 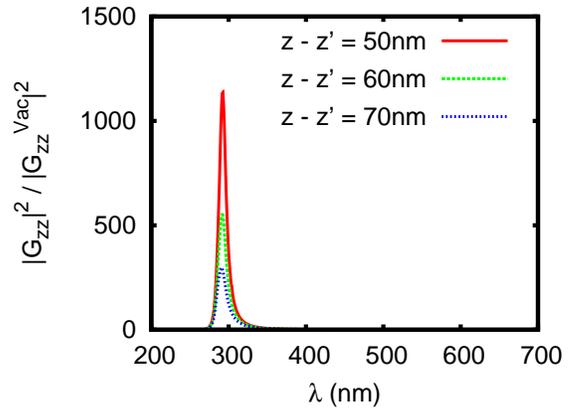, width = 0.45\textwidth}}
  \subfigure[\, Ag, $F_x$]{\epsfig{file = 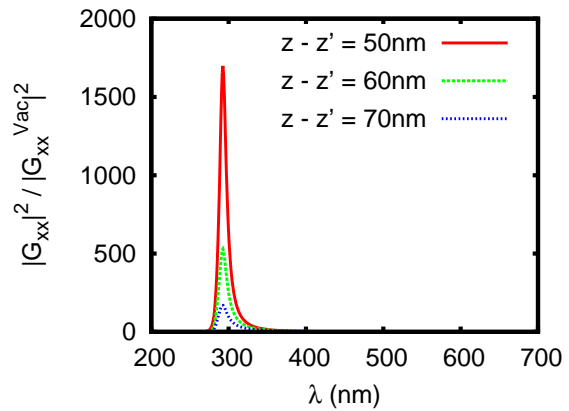 , width = 0.45\textwidth}}
   \caption{Plot of $F_z$ and $F_x$ as function of wavelength for a silver film with thickness $d = 30\,{\rm nm}$. The 
            donor is placed at $z' = -10\,{\rm nm}$ and the position of the accepor is varied such 
            that $z - z' = 50\,{\rm nm}, 60\,{\rm nm}, 70\,{\rm nm}$. \label{Fig:Silverfilm1}}
\end{figure}

That the coupling to the surface plasmons can enhance the energy transfer has been
demonstrated theoretically for different systems like nanoparticles~\cite{GerstenNitzan1984}, 
plasmonic waveguides~\cite{CanoEtAl2010} and films~\cite{BiehsAgarwal2013,Poudel2015} as well as 
graphene~\cite{Velizhanin2012, AgarwalBiehs2013,KaranikolasEtAl2016}. This enhanced energy transfer 
might be exploited for solar energy conversion~\cite{LiEtAl2015}. Experimentally it has been shown 
by Viger et al.~\cite{VigerEtAl2011} and Zhang et al.~\cite{ZhangEtAl2014} that plasmonic nanoparticles 
can enhance the energy transfer rate; Andrew and Barnes~\cite{AndrewBarnes2004} have proven experimentally that 
the F\"{o}rster energy transfer across thin silver films can be enhanced by the coupling to the surface 
plasmon polaritons demonstrating a long-range coupling between the donor-acceptor pair across
films with thicknesses up to 120nm. Although we are here particularly interested in the long-range 
energy transfer across a film, surface plasmons can also be used to mediate the energy transfer
along plasmonic structures as shown theoretically~\cite{CanoEtAl2010,BiehsAgarwal2013,Poudel2015,Velizhanin2012,AgarwalBiehs2013,KaranikolasEtAl2016}.  Recently, Bouchet {\itshape et al.}~\cite{BouchetEtAl2016} have detected this 
long-range energy transfer when both donor and acceptor are placed on a plasmonic platform like a thin 
metal film.

\begin{figure}[Hhbt]
  \subfigure[\, Ag, $F_z$]{\epsfig{file = 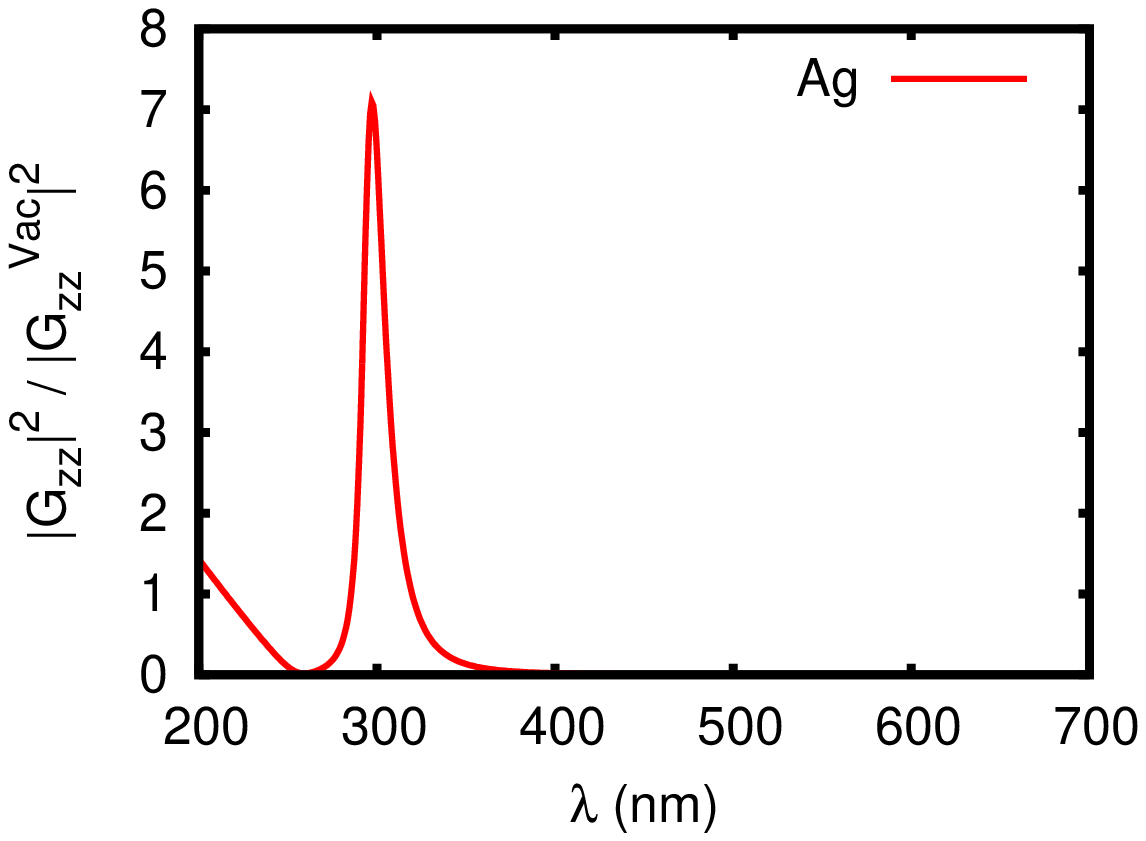, width = 0.45\textwidth}}
  \subfigure[\, Ag, $F_x$]{\epsfig{file = 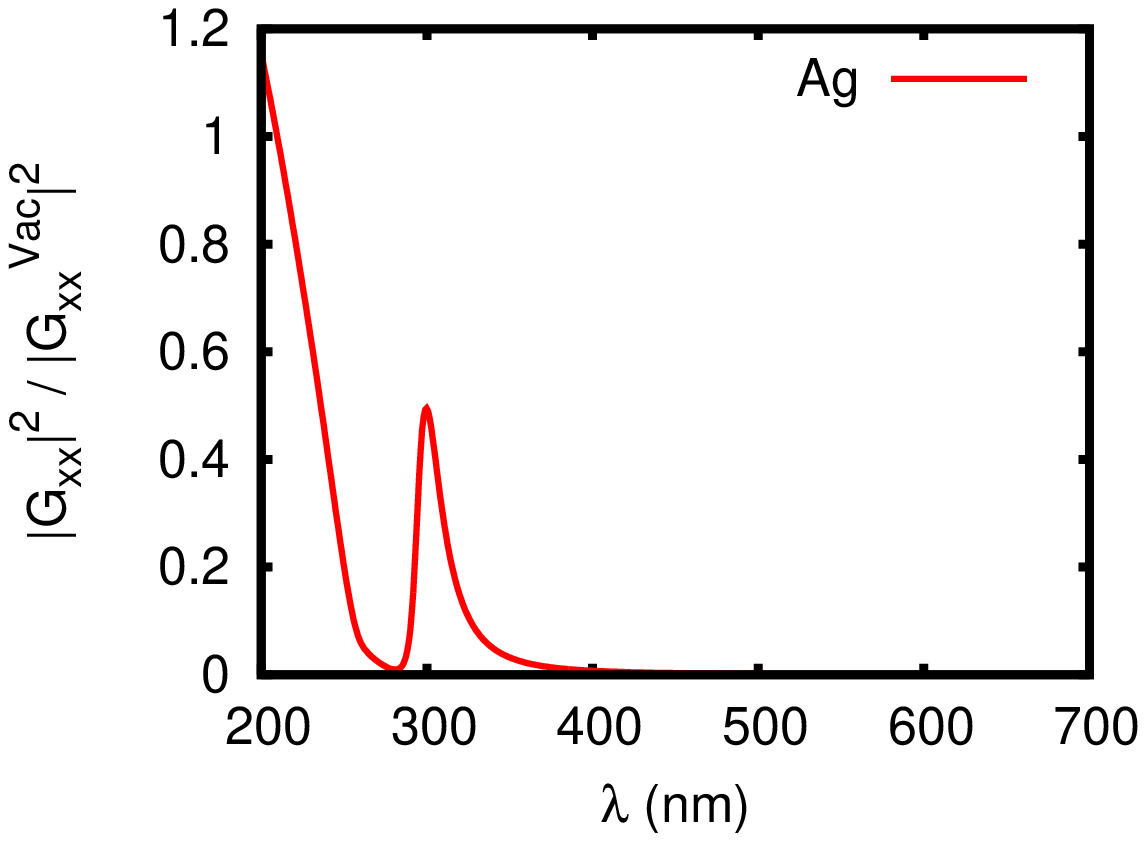 , width = 0.45\textwidth}}
   \caption{Plot of $F_z$ and $F_x$ as function of wavelength. The donor and accepor are placed in a distance of 
            $10\,{\rm nm}$ of each interface (i.e.\ $z' = -10\,{\rm nm}$ and $z - d = 10\,{\rm nm}$) of a silver film with
            thickness $d = 120\,{\rm nm}$. \label{Fig:Silverfilm}        }
\end{figure}

To see how the F\"{o}rster energy transfer is affected by the presence of a metal film, we consider therefore 
first a thin silver layer with permittivity described by the Drude model
\begin{equation}
  \epsilon_{\rm Ag} = \epsilon_\infty - \frac{\omega_\rp^2}{\omega(\omega + \ri \tau^{-1})} 
\label{Eq:Drude}
\end{equation}
with the paramters $\epsilon_\infty = 3.7$, $\omega_\rp = 1.4\cdot10^{16}\,{\rm rad}/{\rm s}$, $\tau = 0.45\cdot10^{-14}\,{\rm s}$. In order to quantify the enhancement of the energy transfer we introduce the enhancement factor
\begin{equation}
  F_{i} \equiv \frac{|\mathds{G}_{ii}|^2}{|\mathds{G}_{ii}^{\rm (vac)}|^2},
\end{equation} 
where $i =x,z$ is the orientation of the dipole-transition matrix element of the donor/acceptor and $\mathds{G}_{ii}^{\rm (vac)}$
is the vacuum Green's function. In Fig.~\ref{Fig:Silverfilm1} we show our results for the enhancement of 
the dipole-dipole interaction due to the presence of a silver film of thickness $d = 30\,{\rm nm}$ with respect 
to the case where this film is replaced by vacuum. It can be seen that the dipole-dipole interaction and therefore 
the F\"{o}rster energy transfer is especially large for $\lambda \approx 300\,{\rm nm}$ where $\epsilon'_{\rm Ag} \approx - 1$ as 
expected. Furthermore it can be seen that at $z - z' = 2d = 60\,{\rm nm}$
no particular effect happens. The F\"{o}rster energy transfer becomes just less important when the distance 
$z-z'$ between the donor-acceptor pair is increased. In Fig.~\ref{Fig:Silverfilm} we show similar results for
a much thicker film with $d = 120\,{\rm nm}$. In this case there can be still seen an enhancement effect for $z$ orientation
but of course the enhancement effect becomes less important when the thickness of the silver film is increased due
the losses inside the metal film. Therefore another kind of material is needed in order to overcome the harmful
effect of the losses. Nonetheless such an enhancement due to the surface plasmons of the silver film have already 
been measured in the experiment by Andrew and Barnes~\cite{AndrewBarnes2004} for film thicknesses of 
$d = 30\,{\rm nm}, 60\,{\rm nm}, 90\,{\rm nm}$ and even $d = 120\,{\rm nm}$. 

%%%%%%%%%%%%%%%%%%%%%%%%%%%%%%%%%%%%%%%%%%%%%%%%%%%%%%%%%%%%%%%%%%%%%%%%%%%%%%%%%%%%
%
% Hyperbolic Metamaterial
%
%%%%%%%%%%%%%%%%%%%%%%%%%%%%%%%%%%%%%%%%%%%%%%%%%%%%%%%%%%%%%%%%%%%%%%%%%%%%%%%%%%%%

\section{Anomalous Transmission for Hyperbolic Materials}

\begin{figure}[Hhbt]
  \epsfig{file = 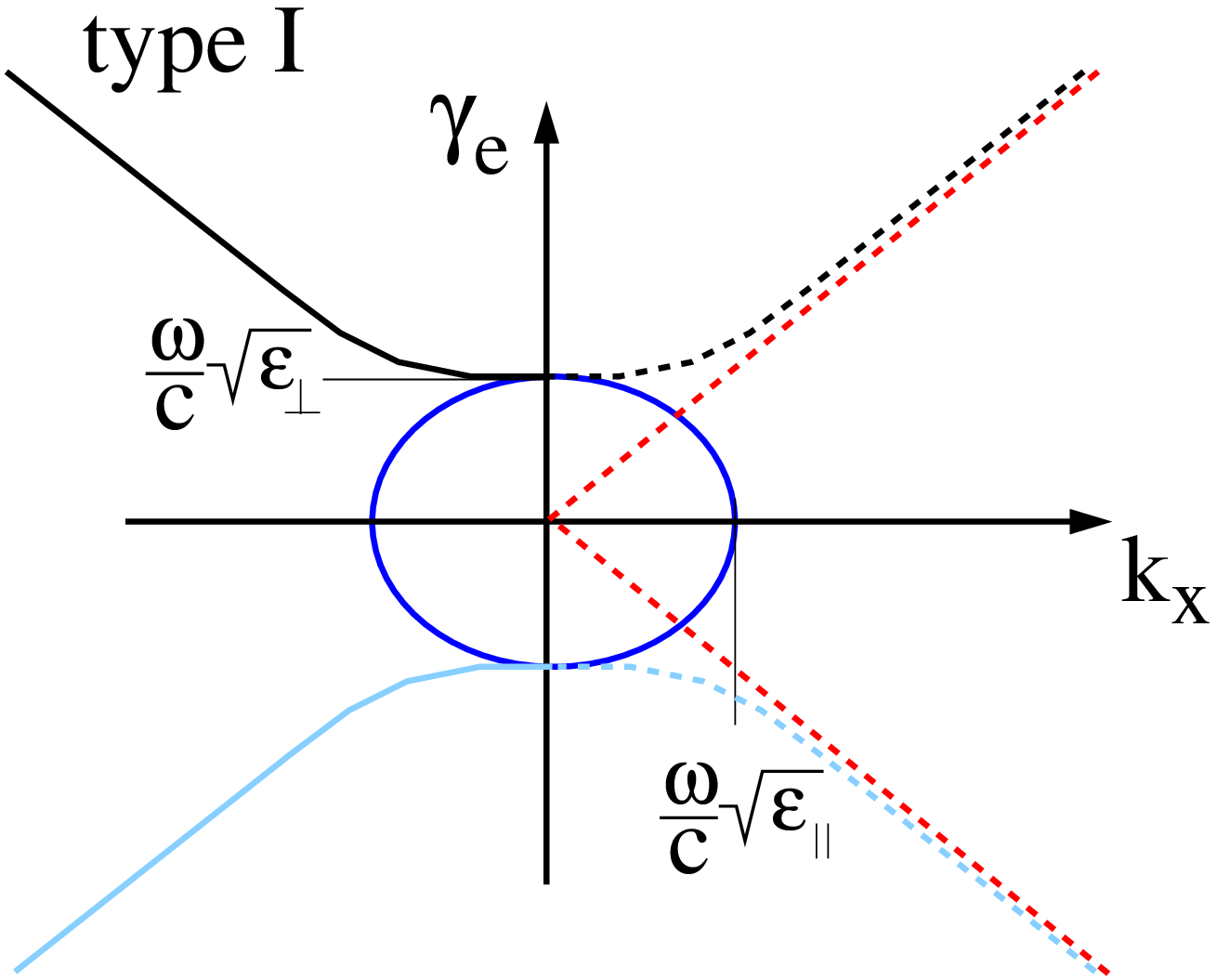, width = 0.4\textwidth}
  \epsfig{file = 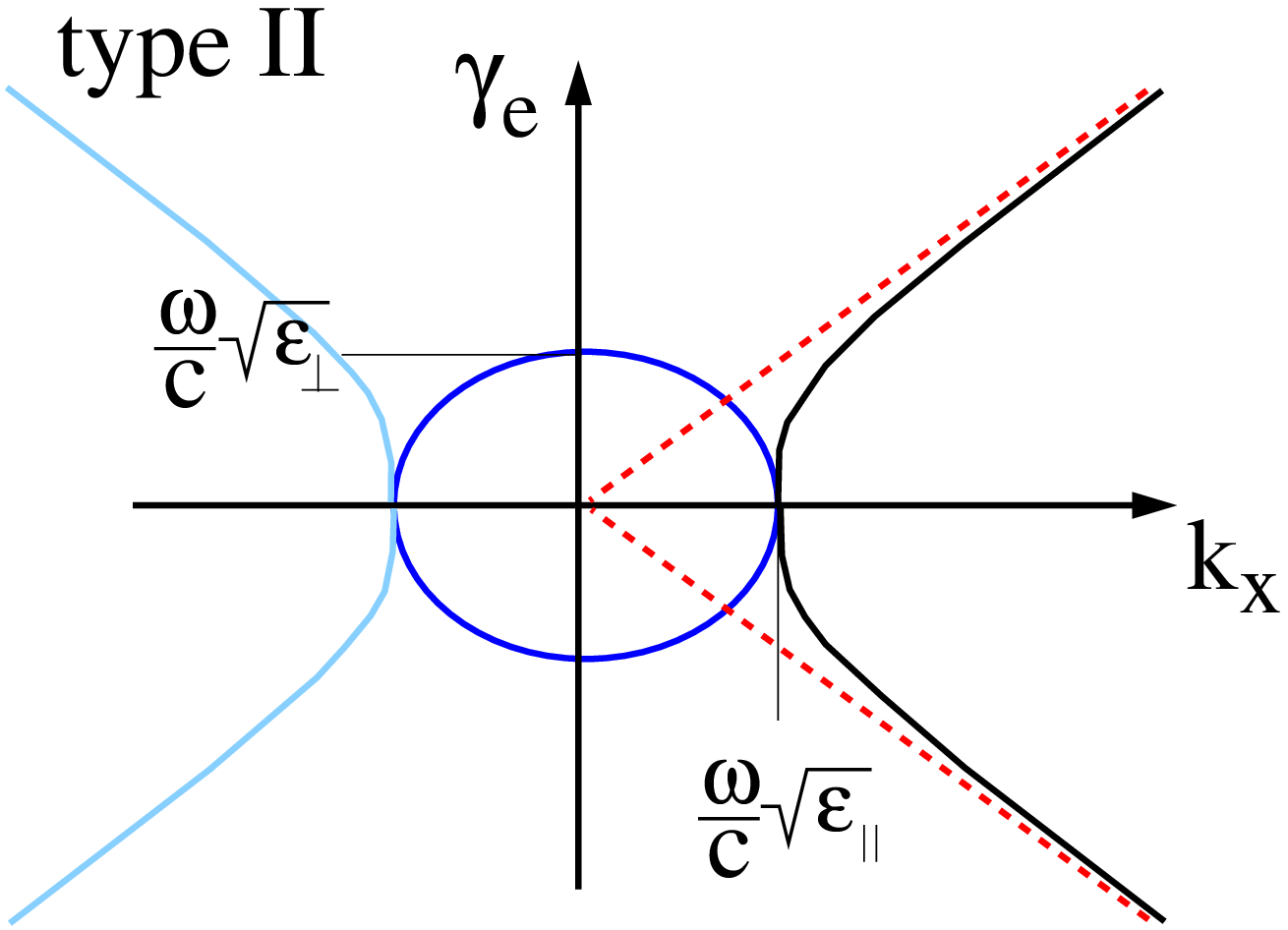, width = 0.4\textwidth}
  \caption{\label{Fig:IsofrequencyLine} Sketch of the isofrequency lines of $\gamma_\re$ for a dielectric uniaxial medium (blue ellipse) and
           a type I/II hyperbolic material (solid black and light blue line). The asymptotes (dashed red lines) are
           given by $k_z = \pm k_x \sqrt{|\epsilon_\perp|/|\epsilon_\parallel|}$.}
\end{figure}

Now, let us replace the ideal left-handed material slab by a hyperbolic or indefinite material~\cite{Smith2003}.
The beauty of such hyperbolic materials is that waves with large kappa ($\kappa \gg k_v$) emitted by a donor 
which would be evanescent in free space are homogenous within these materials. This leads to significant fields on 
the other side of the slab which makes hyperbolic materials very advantageous for energy transfer even for
slabs with an appreciable thickness. Hyperbolic materials are uni-axial materials which do exist in 
nature~\cite{ThompsonEtAl1998,CaldwellEtAl2014,EsslingerEtAl2014,NarimanovKildishev2015,KorzebEtAl2015}. 
But they can also be easily fabricated by combining alternating layers of dielectric and plasmonic materials in 
a periodic structure, for 
instance.  Here, we consider only the non-magnetic case so that the permeability tensor is given by the unit tensor 
and the permittivity tensor is given by ${\rm diag}(\epsilon_\perp, \epsilon_\perp,\epsilon_\parallel)$ 
with respect to the principal axis. In our case the z-axis is the optical axis of the uni-axial material. 
The dispersion relations of the ordinary and extra-ordinary modes inside the uni-axial medium are~\cite{YehBook}
\begin{equation}
  \frac{\gamma_\ro^2}{\epsilon_\perp} + \frac{\kappa^2}{\epsilon_\perp} = k_v^2 \quad \text{and} \quad \frac{\gamma_\re^2}{\epsilon_\perp} +  \frac{\kappa^2}{\epsilon_\parallel} = k_v^2 
\end{equation}
where $\gamma_\ro$ ($\gamma_\re$) is the z-component of the wavevector of the ordinary (extraordinary) mode. Neglecting dissipation 
for a moment we can easily define normal dielectric uni-axial materials as materials with $ \epsilon_\parallel > 0$ and $ \epsilon_\perp > 0$
and hyperbolic materials as materials with $\epsilon_\parallel \epsilon_\perp < 0$, i.e.\ one of both permittivites is positiv
and the other one negative. Due to the property $ \epsilon_\parallel > 0$ and $ \epsilon_\perp > 0$ the iso-frequency curves for dielectric
uni-axial materials in k-space are ellipsoids whereas for hyperbolic materials these iso-frequency curves are hyperboloids. 
For $\epsilon_\parallel < 0$ and $\epsilon_\perp > 0$ --- a type I hyperbolic material --- the iso-frequency curve is a two-sheeted hyperboloid
and for $\epsilon_\parallel > 0$ and $\epsilon_\perp < 0$ --- a type II hyperbolic material --- the iso-freqency curve is a one-sheeted hyperboloid.
In Fig.~\ref{Fig:IsofrequencyLine} the iso-frequency lines of a type I hyperbolic material are sketched in the $k_z$-$k_x$ plane.

%%%%%%%%%%%%%%%%%%%%%%%%%%%%%%%%%%%%%%%%%%%%%%%%%%%%%%%%%%%%%%%%%%%%%%%%%%%%%%%%%%%%
%
% Anomalous transmission across Hyperbolic Metamaterial
%
%%%%%%%%%%%%%%%%%%%%%%%%%%%%%%%%%%%%%%%%%%%%%%%%%%%%%%%%%%%%%%%%%%%%%%%%%%%%%%%%%%%%

Now we will show that similar to the perfect lensing effect the evanescent waves can be amplified inside a hyperbolic slab even in the absence 
of a magnetic response. To this end, we consider the transmission coefficients of a uniaxial slab where the optical axis is normal to the interface, i.e.\
it is along the z-axis. In this case the transmission coefficients for the ordinary and extraordinary modes (which coincide with the s- and p-polarized modes) 
are given by
\begin{align}
  t_\rs &= \frac{4 \gamma_\ro \gamma_v \re^{\ri (\gamma_\ro - \gamma_v) d}}{(\gamma_\ro + \gamma_v)^2 - (\gamma_\ro - \gamma_v)^2 \re^{2 \ri \gamma_\ro d}} \\
  t_\rp &= \frac{4 \epsilon_\perp \gamma_\re \gamma_v \re^{\ri (\gamma_\re - \gamma_v) d}}{(\gamma_\re + \epsilon_\perp \gamma_v)^2 - (\gamma_\re - \epsilon_\perp \gamma_v)^2 \re^{2 \ri \gamma_\re d}}.
\end{align}
The effect of anisotropy is mainly seen in the transmission coefficient $t_\rp$ because only here both permittivities $\epsilon_\perp$ and
$\epsilon_\parallel$ enter. When considering an epsilon-near-zero (ENZ) material with $\epsilon_\perp \rightarrow 0$, then $\gamma_\ro,\gamma_\re \rightarrow 0$ so that the exponential in the denominator of $t_\rs$ and $t_\rp$ is one and we find
\begin{equation}
  t_\rs \rightarrow \re^{- \ri \gamma_v d} \quad \text{and} \quad t_\rp \rightarrow \re^{- \ri \gamma_v d}.
\end{equation} 
That means that similar to the perfect lensing effect, the evanescent waves are amplified by uni-axial material and in particular by 
a hyperbolic structure. Note, that this time the exponential which enhances the evanescent modes stems from the numerator of the transmission 
coefficient which makes this effect more robust against losses. Similarly, when considering an epsilon-near-pole (ENP) material
with $\epsilon_\parallel \rightarrow \infty$, then $t_\rs$ remains unaffected because $\gamma_\ro$ remains unaltered. But for
the extra-ordinary wave we have $\gamma_\re \approx \sqrt{\epsilon_\perp} \omega/c$.
It follows that
\begin{equation}
  t_\rp \rightarrow  A \re^{-\ri \gamma_v d}
\end{equation}
with
\begin{equation}
   A \equiv \frac{4 \epsilon_\perp \sqrt{\epsilon_\perp} k_v \gamma_v \re^{\ri k_v \sqrt{\epsilon_\perp} d}}{ \bigl(k_v \sqrt{\epsilon_\perp} + \epsilon_\perp \gamma_v \bigr)^2 - \bigl( k_v \sqrt{\epsilon_\perp} - \epsilon_\perp \gamma_v \bigr)^2 \re^{2 \ri k_v \sqrt{\epsilon_\perp} d}}. \\
\end{equation} 
Therefore $t_\rp \propto \exp(- \ri \gamma_v d)$ and the evanescent modes are again amplified, if the $\kappa$-dependent prefactor $A$ is not too small.

To summarize, we find that for ENZ and ENP frequencies evanescent waves in vacuum are amplified and the dipole-dipole interaction can 
at least in principle become infinitely large if $z = z' + d$, i.e.\ if the donor and the acceptor are both exactly deposited on the surface 
of the hyperbolic film which cannot be achieved in a real setup. But even if this condition is not perfectly met, we can expect to find an amplified 
energy transfer across a hyperbolic slab at the ENZ and ENP frequencies. Furthermore, from the above derivation suggests that the ENZ 
resonance is more advantageous for transmission than the ENP resonance, because the prefactor fullfills for $\epsilon > 0$ typically $|A| < 1$.

Note, that both regimes of ENZ and ENP were discussed in the context of diffraction suppressed hyperbolic lensing~\cite{FengAndElson2006},
using the canalization regime for hyperbolic lensing~\cite{BelovEtAl2005,BelovHao2006} and directed dipole emission~\cite{CegliaEtAl2014}.
In these cases, the advantage of using ENZ and ENP resonances lies in the resulting very flat iso-frequency line of the extraordinary modes so that the 
group velocity is mainly directed along the optical axis for a very broad band of lateral wavenumbers $\kappa$. This becomes clear when looking at the asymptotes
of the isofrequency lines of $\gamma_\re$ as shown in Fig.~\ref{Fig:IsofrequencyLine}. For large wavenumbers $\kappa$ (evanescent regime) we 
have $\gamma_\re \approx \kappa \sqrt{|\epsilon_\perp|/|\epsilon_\parallel|}$. With this result we can evaluate the ratio of the 
group velocity along the z-axis and the group velocity perpendicular to the z-axis and obtain
\begin{equation}
 \biggl| \frac{\frac{\rd \omega}{\rd \gamma_\re}}{\frac{\rd \omega}{\rd \kappa}} \biggr| = \biggl| \frac{\gamma_\re}{\kappa} \frac{\epsilon_\parallel}{\epsilon_\perp} \biggr| \approx \sqrt{\frac{|\epsilon_\parallel|}{|\epsilon_\perp|}}.
\label{Eq:Forward}
\end{equation}
Hence, at the ENZ and ENP resonance the slope of the isofrequency lines become infinitely small and the group velocity becomes mainly directed 
parallel to the optical axis so that the energy transferred between the donor and acceptor flows preferentially along the line connecting both,
if both are placed along the optical axis. 

%%%%%%%%%%%%%%%%%%%%%%%%%%%%%%%%%%%%%%%%%%%%%%%%%%%%%%%%%%%%%%%%%%%%%%%%%%%%%%%%%%%%
%
% Numerical Example
%
%%%%%%%%%%%%%%%%%%%%%%%%%%%%%%%%%%%%%%%%%%%%%%%%%%%%%%%%%%%%%%%%%%%%%%%%%%%%%%%%%%%%
\section{S-Matrix Calculation of the Dipole-Dipole Interaction}

\begin{figure}[Hhbt]
  \epsfig{file = 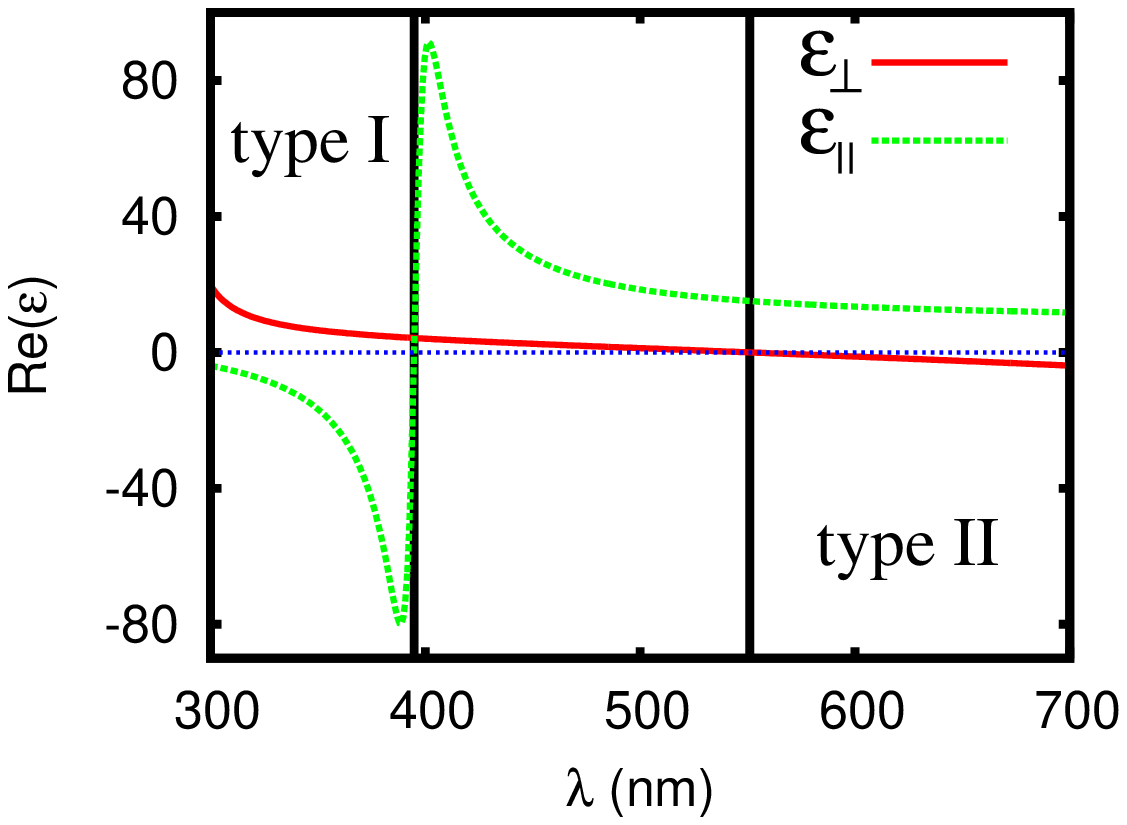, width = 0.4\textwidth}
  \epsfig{file = 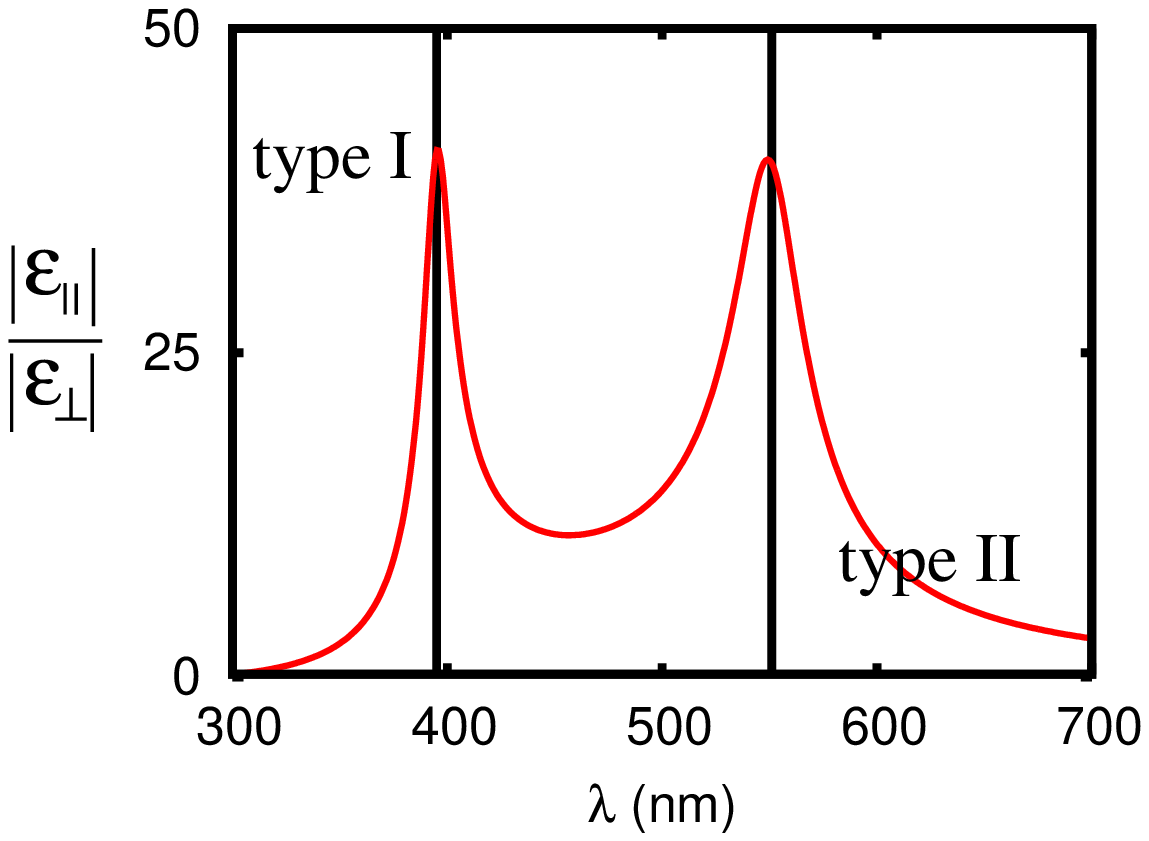, width = 0.4\textwidth}
  \caption{\label{Fig:EpsilonEffective} Top: plot of the real part of the effective permittivites $\epsilon_\perp$ and $\epsilon_\parallel$ for a Ag/TiO$_2$ multilayer hyperbolic material with a silver filling fraction of $f = 0.35$. The vertical lines mark the edges of the hyperbolic bands of type I and type II. Bottom: plot of  $|\epsilon_\perp|/|\epsilon_\parallel|$ manifesting the ENP and ENZ points at $\lambda = 394.8\,{\rm nm}$ and $\lambda = 551.2\,{\rm nm}$.}
\end{figure}

Let us turn to a concrete numerical example. We consider a hyperbolic multilayer metamaterial
made of alternating layers of silver and TiO$_2$ which has for example been used in the experiment 
in Ref.~\cite{KrishnamorthyEtAl2012}. In most treatments of such multilayer materials, 
 effective medium theory (EMT) is used which describes the multilayer structure as a homogenous but
uni-axial material, with an optical axis perpendicular to the interfaces. Within EMT the effective 
permittivities of the multilayer structure can be easily calculated and are given by
\begin{align}
  \epsilon_\perp &= f \epsilon_{\rm Ag} + (1 - f) \epsilon_{{\rm TiO}_2}, \\
  \epsilon_\parallel     &= \frac{\epsilon_{\rm Ag} \epsilon_{{\rm TiO}_2}}{f \epsilon_{{\rm TiO}_2} + (1-f) \epsilon_{\rm Ag} },
\end{align}
where $f$ is the filling fraction of silver and $\epsilon_{\rm Ag}$/$\epsilon_{{\rm TiO}_2}$
are the permittivites of the both constitutents of the multilayer structure. For silver we use the Drude 
model in Eq.~(\ref{Eq:Drude}). TiO$_2$ is transparent in the visible regime. It's 
permittivity $\epsilon_{{\rm TiO}_2}$ is nearly constant in that regime and can be well described 
by the formula~\cite{Devore1951}
\begin{equation}
  \epsilon_{{\rm TiO}_2} = 5.913 + \frac{0.2441}{\lambda^2 - 0.0803}.
\end{equation}
where $\lambda$ is the wavelength in micrometer. The effective permittivties $\epsilon_\perp$ and $\epsilon_\parallel$
are shown in Fig.~\ref{Fig:EpsilonEffective} for a filling fraction of $f = 0.35$. It can be seen that at the edges of the
hyperbolic bands we find the ENP and ENZ points at $\lambda = 394.8\,{\rm nm}$ and $\lambda = 551.2\,{\rm nm}$. To make these 
points more obvious we show in Fig.~\ref{Fig:EpsilonEffective} a plot of $|\epsilon_\perp|/|\epsilon_\parallel|$. By 
changing the filling fraction the position of the ENZ and ENP frequencies can be shifted. When increasing the filling 
fraction the dielectric band in between both hyperbolic bands in Fig.~\ref{Fig:EpsilonEffective} will become smaller 
until $f = 0.5$, then the frequencies of ENP and ENZ coinside since in this case both hyperbolic 
bands (type I and type II) are in juxtaposition with each other. 

However, the EMT is only applicable if the considered wavelengths and distances are much larger than the 
period of the multilayer structure~\cite{KidwaiEtAl2012, IorshEtAl2012,IorshEtAl2012,OrlovEtAl2011,TschikinEtAl2013}. 
Furthermore, EMT cannot account for the ordering of the layers so that it can be very important if we consider 
a finite Ag/TiO$_2$ or TiO$_2$/Ag multilayer material~\cite{TschikinEtAl2013}. Therefore we will use
mainly full S-matrix calculations~\cite{AuslenderHava1996} to determine the transmission coefficients of the multilayer 
hyperbolic film. Here, we choose silver as first layer which is facing the donor. Hence, the last layer facing the
acceptor is made of TiO$_2$. 

\begin{figure}[Hhbt]
  \epsfig{file = 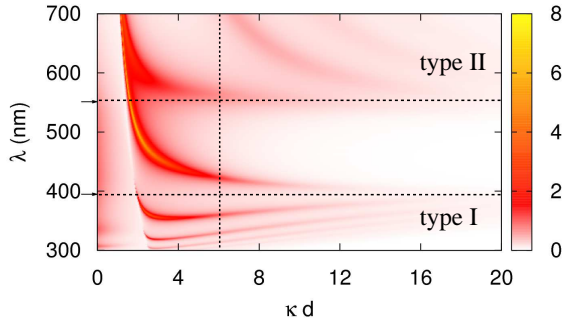, width = 0.45\textwidth}
  \caption{ Plot of the transmission coefficient $|t_\rp \exp(- \kappa d)|^2$ for the hyperbolic Ag/TiO$_2$  multilayer 
            structure $d = 120\,{\rm nm}$ and $N = 24$ with filling fraction of $f = 0.35$. We have multiplied the 
            transmission coefficient with $\exp(- \kappa d)$ to compensate the 
            exponential enhancement of the evanescent waves with large $\kappa \gg \frac{\omega}{c}$. The horizontal lines and arrows
            mark the edges of the hyperbolic bands.
  \label{Fig:TransmissionCoefficient}}
\end{figure}

The transmission coefficient of the extra-ordinary waves $t_\rp$ is plotted in Fig.~\ref{Fig:TransmissionCoefficient}
for a  Ag/TiO$_2$  multilayer structure of thickness $d = 120$ with a filling fraction of $f = 0.35$ of silver. The calculation
was made for $N = 24$ layers, which means that the silver layers have a thickness of 3.5nm and the TiO$_2$ layers have a 
thickness of 6.5nm which is at the lower limit of a realizable structure. In Fig.~\ref{Fig:TransmissionCoefficient} the different
coupled surface modes of the hyperbolic multilayer structure can be seen. Obviously in the type I hyperbolic band the coupled 
surface modes can exhibit negative group velocities which will result in negative refraction~\cite{HoffmanEtAl2007}. Furthermore it can be niceley seen that  all the coupled modes converge for large $\kappa d$ towards the ENZ and ENP frequencies (vertical lines). Therefore at those ENZ and ENP frequencies we have quite a number of surface modes with zero group velocity in direction of the interfaces of the multilayer structure. These are the waves which will contribute dominantly to the F\"{o}rster energy transfer having a group velocity rather along the optical axis (z-direction) than perpendicular to it as shown in Eq.~(\ref{Eq:Forward}). This is so, because for the evanescent waves with large $\kappa \gg k_v$ the exponentials in the Green's function in Eq.~(\ref{Eq:GreensTensor}) introduce a cutoff~\footnote{For the numeric integration we use a cutoff at $\kappa = 300/(z - z')$ which is much larger than the intrinsic cutoff of the integrand at $\kappa \approx 1/(z-z'-d)$.} at $\kappa \approx 1/(z-z'-d)$ at the ENP and ENZ frequencies for the $\kappa$-integral, since $t_\rp \approx \exp(\kappa d)$ in this case. Therefore the major conributions stem from $\kappa$ around $1/(z-z'-d)$. That means in the ideal case $z - z' = d$ all evanescent waves contribute, but in the non-ideal realistic case only evanescent waves up to a finite value of $\kappa \approx 1/(z-z'-d)$ will contribute to the energy transfer.

\begin{figure}[Hhbt]
  \subfigure[\, Ag/TiO$_2$, $F_z$]{\epsfig{file = 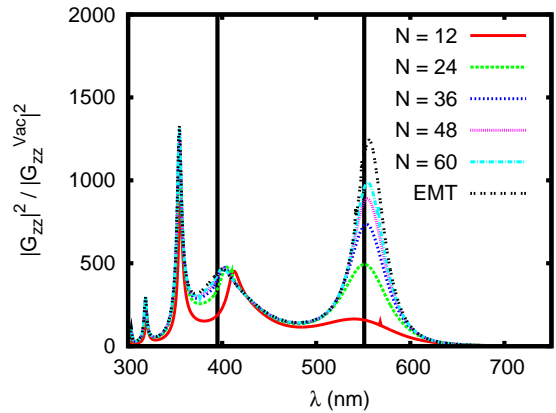, width = 0.45\textwidth}}
  \subfigure[\, Ag/TiO$_2$, $F_x$]{\epsfig{file = 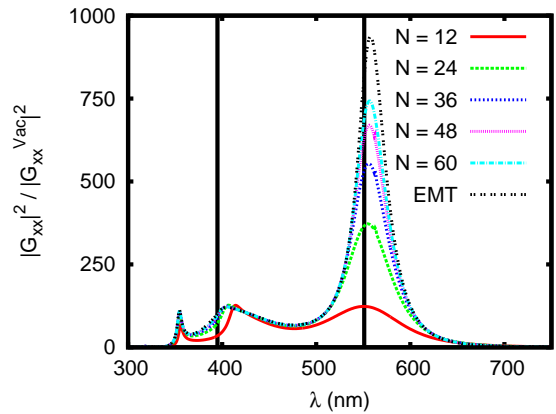, width = 0.45\textwidth}}
  \caption{Plot of $F_z$ and $F_x$ as function of wavelength. Again donor and accepor are placed in a distance of 
            $10\,{\rm nm}$ of each interface (i.e.\ $z' = -10\,{\rm nm}$ and $z - d = 10\,{\rm nm}$) of the film of thickness  $d = 120\,{\rm nm}$. 
            Here the film is given by a Ag/TiO$_2$ multilayer structure with $N = 12,24,36,48,60$ layers. The filling fraction of silver is $f = 0.35$.
            For comparision the exact and the EMT results are shown. The vertical lines correspond again to the edges of the hyperbolic bands
            as in Fig.~\ref{Fig:EpsilonEffective}. \label{Fig:ExactEffective}}
\end{figure}

In Fig.~\ref{Fig:ExactEffective} we show now the results for the enhancement factors $F_x$ and $F_z$ of the F\"{o}rster energy transfer 
by a Ag/TiO$_2$ hyperbolic multilayer structure. It can be seen that the enhancement is especially large close to the ENZ and ENP 
frequencies. Furthermore, the enhancement is more than two orders of magnitude larger than for a single silver film in Fig.~\ref{Fig:Silverfilm}, 
so that this enhancement of energy transfer due to the ENP and ENZ resonances is much larger than the enhancement due 
to the thin film surface plasmons. Since the latter has already been measured by Andrew and Barnes~\cite{AndrewBarnes2004} the ENP and ENZ enhancement 
should be easily measurable. Note, that by increasing the number of layer $N$ in the multilayer film or by increasing the distance of the 
donor-acceptor pair with respect to the film the exact S-matrix result converges to the EMT result. 

\begin{figure}[Hhbt]
  \epsfig{file = 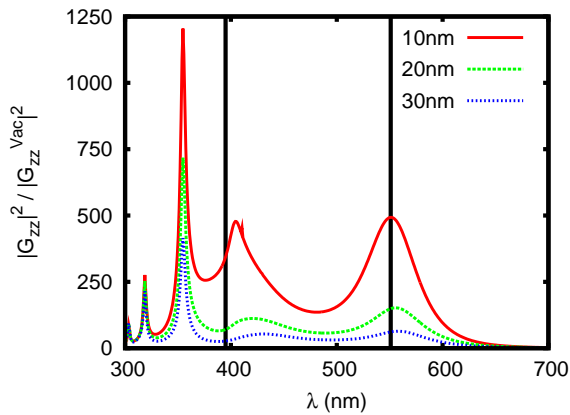, width = 0.45\textwidth}
  \caption{Plot of $F_z$ for a Ag/TiO$_2$ multilayer structure with $N = 24$ and different donor-acceptor positions. 
           The positions are given by $z' = -10\,{\rm nm}, -20\,{\rm nm}, -30\,{\rm nm}$ with 
           corresponding $z - d = 10\,{\rm nm}, 20\,{\rm nm}, 30\,{\rm nm}$.\label{Fig:DiffDistances}}
\end{figure}

\begin{figure}[Hhbt]
  \subfigure[\, Ag/TiO$_2$, $F_z$]{\epsfig{file = 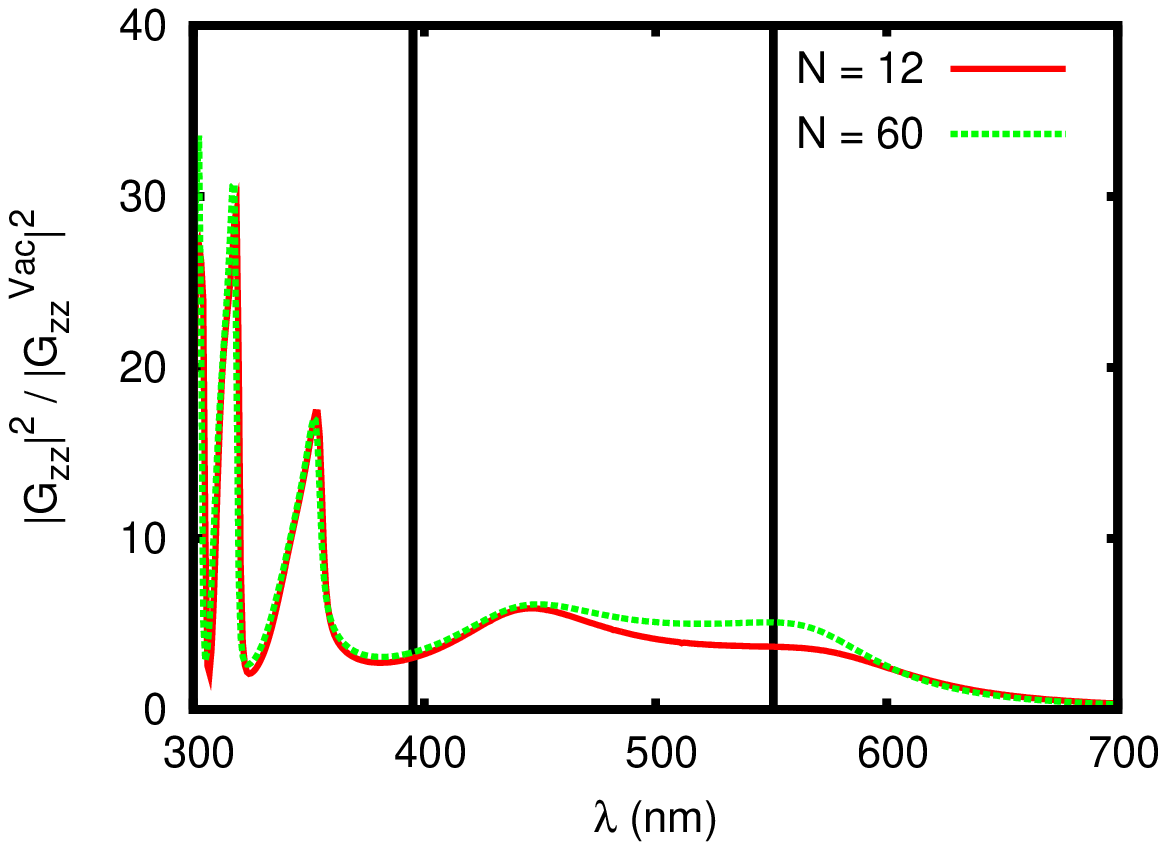, width = 0.45\textwidth}}
  \subfigure[\, Ag/TiO$_2$, $F_x$]{\epsfig{file = 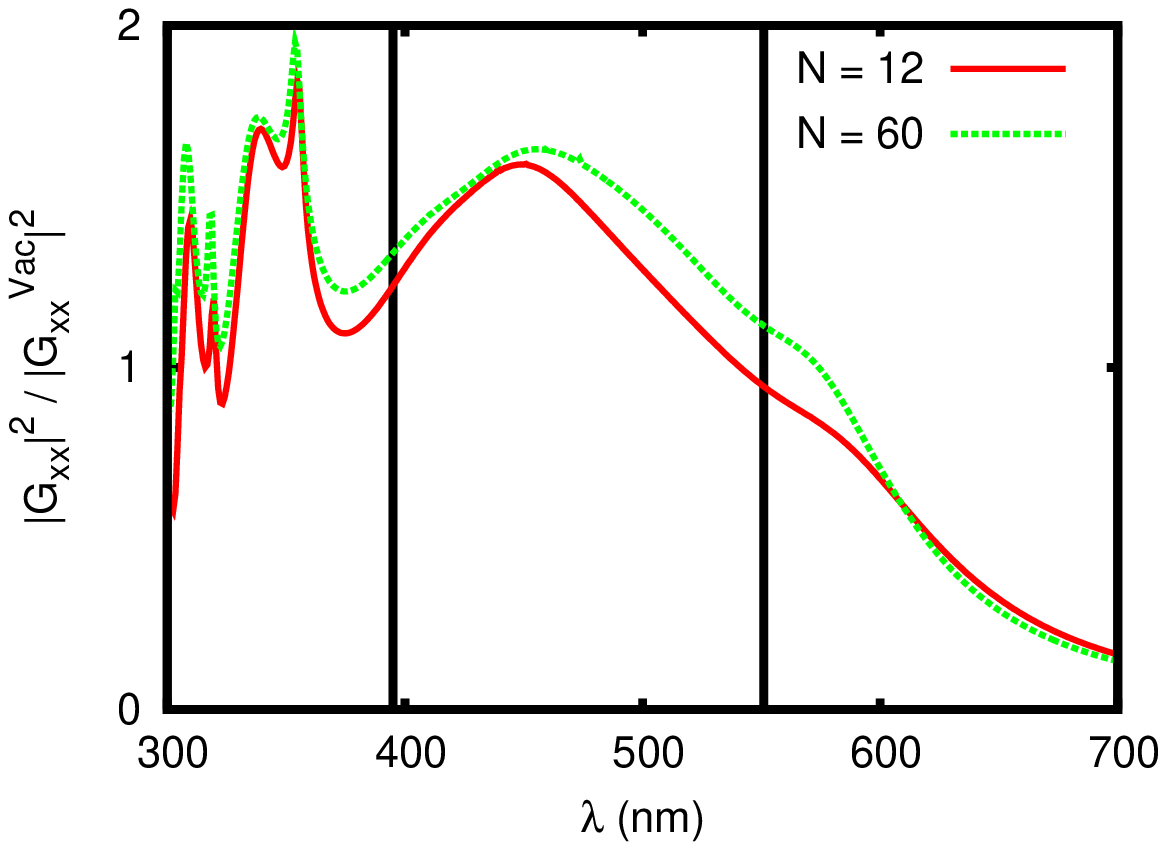, width = 0.45\textwidth}}
  \caption{ Same as in Fig.~\ref{Fig:ExactEffective} but with $z' = -100\,{\rm nm}$ and $z = d + 100\,{\rm nm}$ for $N = 12, 60$. The EMT 
            result is not shown because it practically coinsides with the result for $N = 60$. \label{Fig:100nm}}
\end{figure}

The position of the peaks can be explained be the fact that the dominant contributions to the energy transfer stem 
from $\kappa d \approx d/(z-z'-d) = 6$ in this case. The vertical line in Fig.~\ref{Fig:TransmissionCoefficient} is exactly at this value. 
It can be seen that the frequencies at which the coupled surface modes cross this vertical line coincide with the resonances of the 
energy transmission in Fig.~\ref{Fig:ExactEffective}. If we would increase the donor-acceptor distance $z - z'$ the vertical line 
would move to smaller $\kappa$ values in Fig.~\ref{Fig:TransmissionCoefficient} so that the resonances will shift accordingly. 
This shifting of the resonances is shown in Fig.~\ref{Fig:DiffDistances}. Obviously, the enhancement around the ENZ and ENP frequencies
gets smaller and smaller, when the distance of the donor and acceptor with respect to the surface is increased. However as
can be seen in Fig.~\ref{Fig:100nm}, there can still be an enhancement of 30 for the energy transfer inside the type I hyperbolic 
band if the distance of the donor and acceptor with respect to the surface of the hyperbolic medium has relative large values as 
for example $100\,{\rm nm}$ so that $z - z' = 320\,{\rm nm}$.

%%%%%%%%%%%%%%%%%%%%%%%%%%%%%%%%%%%%%%%%%%%%%%%%%%%%%%%%%%%%%%%%%%%%%%%%%%%%%%%%%%%%
%
% Conclusion
%
%%%%%%%%%%%%%%%%%%%%%%%%%%%%%%%%%%%%%%%%%%%%%%%%%%%%%%%%%%%%%%%%%%%%%%%%%%%%%%%%%%%%

\section{Conclusion}

In conclusion, we have discussed the perfect lens effect and its pendant for hyperbolic
metamaterials in the context of F\"{o}rster energy transfer between a donor-acceptor
pair placed on each side of the perfect or hyperbolic lens. We have demonstrated that in 
principle in both cases one can have an infinitely large dipole-dipole interaction for 
precisely defined positions of the donor-acceptor pair and well-defined frequencies
which coincide with the surface plasmon frequency in the case of a perfect lens and
the epsilon-near zero or epsilon-near infinity frequency for the hyperbolic lens. Due to 
losses this interactions becomes finite but it remains large compared to the 
dipole-dipole interaction in vacuum. The main reason for such large enhancements is that
the enhancment effect in HMM is much more robust against the negative influence of losses,
whereas the perfect lens effect is quite susceptible to losses.
Finally, we have studied the effect numerically
using an exact S-matrix method and realistic material parameters for a Ag/TiO$_2$
mulilayer HMM. The numerical results show that FRET can be enhanced by two to three 
orders of magnitude by using a HMM instead of a thin silver film of same thickness. 
Since the plasmonic enhancement of FRET for a thin silver film has already been observed,
a measurement of the enhancment due to the hyperbolic material should be feasible.
Such a realization is important not only for FRET but for other areas like production 
of quantum entanglement~\cite{AgarwalBook}.

%%%%%%%%%%%%%%%%%%%%%%%%%%%%%%%%%%%%%%%%%%%%%%%%%%%%%%%%%%%%%%%%%%%%%%%%%%%%%%%%%%%%
%
% Acknowledgements
%
%%%%%%%%%%%%%%%%%%%%%%%%%%%%%%%%%%%%%%%%%%%%%%%%%%%%%%%%%%%%%%%%%%%%%%%%%%%%%%%%%%%% 

%\begin{acknowledgments} 
%\end{acknowledgments}  

\bibliographystyle{plain}

\begin{thebibliography}{30}  
  % VdW
  \bibitem{MilonniBook} P. W. Milonni, {\em The Quantum Vacuum}, (Academic Press, 1994).
  \bibitem{Volokitin2007} A. I. Volokitin and B. N. J. Persson, Rev. Mod. Phys. {\bf 79}, 1291 (2007).
  % Foerter Transfer
  \bibitem{Forster} T. F\"{o}rster, Ann. Phys. {\bf 2}, 55 (1948).
  \bibitem{DungEtAl2002} H. T. Dung, L. Kn\"{o}ll, and D.-G. Welsch, Phys. Rev. A {\bf 65}, 043813 (2002).
  % Heat transfer nanoparticles
  \bibitem{BiehsAgarwal2013b} S.-A. Biehs and G. S. Agarwal, J. Opt. Soc. Am. B {\bf 30}, 700 (2013). 
   %QI protocols
   \bibitem{Nielsen} M. A. Nielsen and I. L. Chuang, {\em Quantum Computation and Quantum Information}, (Cambridge University Press, Cambridge,2000).
   \bibitem{Bouchoule2002} I. Bouchoule and K. Molmer, Phys. Rev. A {\bf 65}, 041803 (2002).
   \bibitem{Isenhower2010} L. Isenhower, E. Urban, X. L. Zhang, A. T. Gill, T. Henage, T. A. Johnson, T. G. Walker, and M. Saffman, Phys. Rev. Lett. {\bf 104}, 010503 (2010).
  % PAIRWISE EXCITATION
  \bibitem{VaradaGSA1992} G. V. Varada and G. S. Agarwal, Phys. Rev. A {\bf 45}, 6721 (1992).
  \bibitem{HaakhEtAl2015} H. R. Haakh and D. Martin-Cano, ACS Photonics {\bf 2}, 1686 (2015).
  \bibitem{HettichEtAl2002} C. Hettich, C. Schmitt, J. Zitzmann, S. K\"{u}hn, I. Gerhardt and V. Sandoghdar, Science {\bf 298}, 385 (2002).
  % Rydberg blockade
  \bibitem{SaffmanEtAl} M. Saffman, T. G. Walker, T. G. and K. Molmer, Rev. Mod. Phys. {\bf 82},  2313 (2010).
  \bibitem{GilletEtAl} J. Gillet, G. S. Agarwal and T. Bastin, Phys. Rev. A {\bf 81}, 013837 (2010).

   %Large DD interaction 
   \bibitem{GSASub} G. S. Agarwal and S. Dutta Gupta, arXiv:quant-ph/0011098.
   \bibitem{ElGanainy2013} R. El-Ganainy, S. John, New J. Phys. {\bf 15}, 083033 (2013).
     %Forster measurement
  \bibitem{AndrewBarnes2004} P. Andrew and W. L. Barnes, Science {\bf 306}, 1002 (2004).
  % FRET    
  \bibitem{Velizhanin2012} K. A. Velizhanin and T. V. Shahbazayan, Phys. Rev. B {\bf 86}, 245432 (2012).
  \bibitem{AgarwalBiehs2013} G. S. Agarwal and S.-A. Biehs, Opt. Lett. {\bf 38}, 4421 (2013).
  \bibitem{KaranikolasEtAl2016} V. D. Karanikolas,  C. A. Marocico, and A. L. Bradley, Phys. Rev. B {\bf 93}, 035426 (2016).
  \bibitem{BiehsAgarwal2013} S.-A. Biehs and G. S. Agarwal, Appl. Phys. Lett. {\bf 103}, 243112 (2013).
  \bibitem{Poudel2015} A. Poudel, X. Chen, and M. A. Ratner, arxic:1601.04338v1 (2015).
   \bibitem{LiEtAl2015} J. Li, S. K. Cushing, F. Meng, T. R. Senty, A. D. Bristow, and	N. Wu, Nat. Phot. {\bf 9}, 601 (2015).
  \bibitem{CanoEtAl2010} D. Martin-Cano, L. Martın-Moreno, F.J. Garcıa-Vidal, and E. Moreno, Nano Lett. {\bf 10}, 3129(2010).
  \bibitem{BouchetEtAl2016} D. Bouchet, D. Cao, R. Carminati, Y. De Wilde, and V. Krachmalnicoff, Phys. Rev. Lett. {\bf 116}, 037401 (2016).
  % uni-axial left-handed material
  \bibitem{HuChui2002} L. Hu and S. T. Chui, Phys. Rev. B {\bf 66}, 085108 (2002).
  % indefinite Material
  \bibitem{Smith2003} D. R. Smith and D. Schurig,  Phys. Rev. Lett. {\bf 90}, 077405 (2003).
  % large LDOS
  \bibitem{SmolyaninovNarimanov2010} I. I. Smolyaninov and E. E. Narimanov, Phys. Rev. Lett. {\bf 105}, 067402 (2010).
  % spontaneous emission  
  \bibitem{ZubinEtAl2012} Z. Jacob, I. Smolyaninov, and E. Narimanov, Appl. Phys. Lett. {\bf 100}, 181105 (2012).
  \bibitem{PoddubnyEtAl2011} A. N. Poddubny, P. A. Belov, and Y. S. Kivshar, Phys. Rev. A {\bf 84}, 023807 (2011).
  \bibitem{ChebykinEtAl2012} A. V. Chebykin, A. A. Orlov, and P. A. Belov, Opt. Spectr. {\bf 109}, 938 (2010).
  \bibitem{PotemkinEtAl2012} A. S. Potemkin, A. N. Poddubny, P. A. Belov, and Y. S. Kivshar, Phys. Rev. A {\bf 86}, 023848 (2012).
  \bibitem{KidwaiEtAl2012} O. Kidwai, S. V. Zhukovsky, and J. E. Sipe, Phys. Rev. A {\bf 85}, 053842 (2012).
  \bibitem{IorshEtAl2012} I. Iorsh, A. N. Poddubny, A. A. Orlov, P. A. Belov, and Y. S. Kivshar, Phys. Lett. A {\bf 376}, 185 (2012)
  \bibitem{KidwaiEtAl2011} O. Kidwai, S. V. Zhukovsky, and J. E. Sipe, Opt. Lett. {\bf 36}, 2530 (2011).
  \bibitem{OrlovEtAl2011} A. A. Orlov, P. M. Voroshilov, P. A. Belov, and Y. S. Kivshar, Phys. Rev. B {\bf 84}, 045424 (2011).
  % topological transition (experiments)  
  \bibitem{KimEtAl2012} J. Kim V. P. Drachev, Z. Jacob, G. V. Naik, A. Boltasseva, E. E. Narimanov, and V. M. Shalaev, Opt. Exp. {\bf 20}, 8100 (2012).
  \bibitem{KrishnamorthyEtAl2012} H. N. S. Krishnamoorthy, Z. Jacob, E. E. Narimanov, I. Kretzschmar, and V. M. Menon, Science {\bf 336}, 205 (2012).
  \bibitem{GalfskyEtAl2015} T. Galfsky, H. N. S. Krishnamoorthy, W. Newman, E. E. Narimanov, Z. Jacob, and V. M. Menon, Optica {\bf 2}, 62 (2015).
  \bibitem{WangEtAl2015} Y. Wang, H. Sugimoto, S. Inampudi, A. Capretti, M. Fuji, and L. Dal Negro, Appl. Phys. Lett. {\bf 106}, 241105 (2015).
  % hyperlens
  \bibitem{JacobEtAl2006} Z. Jacob, L. V. Alekseyev, E. Narimanov, Opt. Exp. {\bf 14}, 8247 (2006).
  % diffraction suppressed hyperbolic lensing 
   \bibitem{FengAndElson2006} S. Feng and J. M. Elson, Opt. Exp. {\bf 14}, 216 (2006).
   \bibitem{CegliaEtAl2014} D. de Ceglia, M. A. Vincenti, S. Campione, F. Capolino, J. W. Haus, and M. Scalora, Phys. Rev. B {\bf 89}, 075123 (2014).
  % negative refraction 
  \bibitem{SmithEtAl2004} D. R. Smith, P. Kolinko, and D. Schurig, J. Opt. Soc. Am. B {\bf 21}, 1032 (2004).
  \bibitem{HoffmanEtAl2007} A. J. Hoffman, L. Alekseyev, S. S. Howard, K. J. Franz, D. Wasserman, V. A. Podolskiy, E. E. Narimanov, S. L. Sivco, and C. Gmachl, Nature Mat. {\bf 6}, 946 (2007).
  % thermal radiation 
  \bibitem{Nefedov2011} I. S. Nefedov and C. R. Simovski, Phys. Rev. B {\bf 84}, 195459 (2011).
  \bibitem{Biehs2012} S.-A. Biehs, M. Tschikin, P. Ben-Abdallah, Phys. Rev. Lett.  {\bf 109}, 104301  (2012).
  \bibitem{GuoEtAl2012} Y. Guo, C. L. Cortes, S. Molesky, and Z. Jacob, Appl. Phys. Lett. {\bf 101}, 131106 (2012).
  \bibitem{Biehs2} S.-A. Biehs, M. Tschikin, R. Messina, and P. Ben-Abdallah, Appl. Phys. Lett.  {\bf 102} 131106 (2013).
  \bibitem{ShiEtAl2015} J. Shi, B. Liu, P. Li, L. Y. Ng, and S. Shen, Nano Letters {\bf 15}, 1217 (2015).
  \bibitem{Biehs2015} S.-A. Biehs, S. Lang, A. Yu. Petrov, M. Eich, and P. Ben-Abdallah, Phys. Rev. Lett. {\bf 115}, 174301 (2015).
  % natural hyperbolic materials
  \bibitem{ThompsonEtAl1998} D. W. Thompson, M. J. DeVries, T. E. Tiwald, and J. A. Woollam, Thin Solid Films {\bf 313-314}, 341 (1998).
  \bibitem{CaldwellEtAl2014} J. D. Caldwell, A. Kretinin, Y. Chen, V. Giannini, M. M. Fogler, Y. Francescato, C. T. Ellis, J. G. Tischler, C. R. Woods, A. J. Giles, M. Hong, K. Watanabe, T. Taniguchi, S. A. Maier, K. S. Novoselov, arXiv:1404.0494.
  \bibitem{EsslingerEtAl2014} M. Esslinger, R. Vogelgesang, N. Talebi, W. Khunsin, P. Gehring, S. de Zuani, B. Gompf, and K. Kern, ACS Photonics, {\bf 1}, 1285(2014).
  \bibitem{NarimanovKildishev2015} E. E. Narimanov and A. V. Kildishev, Nature Photonics {\bf 9}. 214 (2015).
  \bibitem{KorzebEtAl2015}  K. Korzeb, M. Gajc, and D. A. Pawlak, Opt. Expr. {\bf23}, 25406 (2015).
  % ENZ
  \bibitem{Fleury2013} R. Fleury and A. Al\'{u}, Phys. Rev. B {\bf 87}, 201101(R) (2013).
  \bibitem{Silvereinha2006} M. Silveirinha and N. Engheta, Phys. Rev. Lett. {\bf 97}, 157403 (2006).
  \bibitem{EdwardsEtAl2008} B. Edwards, A. Al\'{u}, M. E. Young, M. Silveirinha, and N. Engheta, Phys. Rev. Lett. {\bf 100}, 033903 (2008).
  % Perfect Lens
  \bibitem{Pendry2000} J. B. Pendry, Phys. Rev. Lett. {\bf 85}, 3966 (2000).
  \bibitem{GerstenNitzan1984} J. I. Gersten and A. Nitzan, Chem. Phys. Lett. {\bf 104}, 31 (1984).
  %FRET
  \bibitem{VigerEtAl2011} M. L. Viger, D. Brouard, and D. Boudreau, J. Phys. Chem. C {\bf 115}, 2974 (2011).
  \bibitem{ZhangEtAl2014} X. Zhang, C. A. Marocico, M. Lunz, V. A. Gerard, Y. K. Gunko, V. Lesnyak, N. Gaponik, A. S. Susha,A. L. Rogach, and A. L. Bradley†, ACS Nano {\bf 8}, 1273 (2014).  
  
  % Yehbook  
  \bibitem{YehBook} P. Yeh, {\itshape Optical Waves in Layered Media}, (John Wiley \& Sons, New Jersey, 2005).
 % canalisation regime
  \bibitem{BelovEtAl2005} P. A. Belov, C. R. Simovski, and P. Ikonen, Phys. Rev. B {\bf 71}, 193105 (2005).
  \bibitem{BelovHao2006} P. A. Belov and Y. Hao, Phys. Rev. B {\bf 73}, 113110 (2006).
  %Limitations of EMT
   \bibitem{TschikinEtAl2013} M. Tschikin, S.-A. Biehs, R. Messina, P. Ben-Abdallah, J. Opt. {\bf 15}, 105101 (2013).
  % S-Matrix
   \bibitem{AuslenderHava1996} M. Auslender and S. Have, Opt. Lett. {\bf 21}, 1765 (1996).
  % TiO2
  \bibitem{Devore1951} J. R. Devore, J. Opt. Soc. Am. {\bf 41}, 416 (1951).
  \bibitem{AgarwalBook} See Sec.~15.3 of G. S. Agarwal, {\em Quantum Optics}, (Cambridge University Press, Cambridge, 2012).
\end{thebibliography}

\end{document}